\documentstyle[11pt,mrs2003,epsfig]{article}
\begin{document} 
\title{Brane World Moduli and the CMB} 
\author{Ph. Brax${}^1$, C. van de Bruck${}^2$, A.--C. Davis${}^{3}$ 
and C.S. Rhodes$^3$}
\affil{$^1$Service de Physique Th\'eorique, CEA-Saclay\\ 
F-91191, Gif/Yvette cedex, France \\
$^2$Astrophysics Department, Oxford University, Keble Road \\
Oxford OX1 3RH, U. K.
\\
$^3$Department of Applied Mathematics and Theoretical Physics, 
Center for Mathematical Sciences,\\
University of Cambridge, Wilberforce Road, Cambridge CB3 0WA, U.K.} 
 
\begin{abstract} 
The evolution of moduli fields, which naturally appear in higher 
dimensional models such as brane worlds, and their effects on the 
anisotropies of the cosmological microwave background radiation 
is discussed. 
\end{abstract} 
 
\section{Introduction} 
Extra spatial dimensions appear naturally in models, which aim to 
combine the principles of Quantum Mechanics and General Relativity. 
Among the possible models, brane world scenarios have attracted 
a lot of attention recently \cite{review}. The cosmological evolution of 
brane worlds are of considerable interest, because some scenarios predict 
observable consequences, such as the variation of the fine structure 
constant, the gravitational coupling or masses of particles. This is
due to the existence of massless scalar degrees of freedom 
(moduli fields), which couple to the matter sector in the theory.
Variations of masses or Newton's 
constant have an important impact in cosmology. 
In the following, the effects of moduli fields appearing in brane 
world scenarios on the anisotropies in the cosmic microwave background 
radiation are summarized (details can be found in \cite{rhodes}). 
As we will show, cosmological observations put 
stringent constraints on parameters in the low energy effective theory. 
As such, cosmological observations can be used complementarily to local 
experiments, i.e. experiments in the solar system, in order to constrain 
the coupling of scalar degrees of freedom to matter. 

\section{The low energy effective action}
We begin by summarizing the properties of the 
low energy effective action of a two brane system 
with a bulk scalar field. At low energy, there are two scalar 
degrees of freedom. Their higher--dimensional interpretation is 
as follows: first, there is the bulk scalar field, which propagates 
in all spatial directions. The zero mode, i.e. the massless excitation 
of the bulk scalar field is one degree of freedom in the low energy 
effective action. The other scalar degree of freedom in the low energy 
effective action is the physical distance between the branes. These two
scalar degrees of freedom generally evolve during the cosmological 
evolution. Apart from the two scalar degrees of freedom, there might 
be some forms of matter, confined on the individual branes.

There are different ways of obtaining the low energy effective action, 
see e.g. \cite{kannosoda} and \cite{brax}. Here, we write down the effective 
action in the Einstein frame, obtained with the help of the 
moduli space approximation in \cite{brax}:
\begin{eqnarray}\label{fundamental}
S_{\rm EF} &=& \frac{1}{16\pi G_N} 
\int d^4x \sqrt{-g}\left[ {\cal R} -  \frac{12 \alpha^2}{1+2\alpha^2}
(\partial \varphi)^2 - \frac{6}{2\alpha^2 + 1}(\partial R)^2 
- V(\varphi,R)\right],
\end{eqnarray}
and the matter action:
\begin{equation}
S_{\rm Matter} = 
S_{{\rm Matter},1}(\psi_1, A(\varphi,R)^2 g_{\mu\nu}) + 
S_{{\rm Matter},2}(\psi_2, B(\varphi,R)^2 g_{\mu\nu}),
\end{equation}
The moduli fields $\varphi$ and $R$ are non--trivial combinations of the 
bulk scalar field zero mode and the physical distance between the 
branes. The functions $A(\varphi,R)$ and $B(\varphi,R)$ describe the 
coupling of the two moduli fields to the matter fields confined on each 
branes. Because of the warping of the extra dimension, the functions 
$A$ and $B$ are generally different, i.e. matter on the individual branes
couple differently to the matter on the branes. There is only one free 
parameter in the low energy effective theory which we will denote by 
$\alpha$. This parameter has to be small in order for the theory to be 
consistent with local experiments \cite{brax}. For generality, we have included a 
potential energy, $V(\varphi,R)$, for the fields. 

\section{Moduli Evolution and the CMB}
Solving the field equations derived from the action above, shows that 
the field $R$ decays, i.e. it approaches zero in the matter dominated 
era. This is a valuable feature of the model as local experiments demand 
that the field value today has to be small \cite{brax}. 

Both fields evolve during the cosmological evolution and each will contribute
to the expansion rate of the universe and the evolution of perturbations within 
the universe. We have studied the effect on the evolution of perturbations 
of each individual field \cite{rhodes}, discussing one field at each time only. 
The cosmological parameter are chosen such that the cosmological model today 
is the $\Lambda$CDM model with $\Omega_{\rm CDM}=0.3$, 
$\Omega_{\Lambda} = 0.7$ and $h=0.7$. 

We begin first with the field $\varphi$, i.e. the field with constant coupling. 

\subsection{The case of constant coupling}
In this case, the coupling of the field to matter is determined by the 
free parameter $\alpha$. In Figure 1 we plot the unnormalized power spectra 
for different values of $\alpha$. As it can be seen, both the amplitude as 
well as the positions of all peaks are affected. The case for $\alpha=0$ 
corresponds to the $\Lambda$CDM model based on General Relativity.

It can be shown that the field contributes to the distance to the last 
scattering surface as well as to the scaling of the dark matter density 
\cite{brax}: whereas in the standard model $\rho_{\rm CDM} \propto a^{-3}$, 
where $a$ is the scale factor, this no longer true in the case when 
$\alpha$ is not zero. Thus, at the time of last scattering the matter 
content is {\it different} from the standard case. As known, the 
matter content determines both the positions and the amplitude of the 
peaks \cite{dodelson}. We also found that the coupling has an impact on the 
magnitude on the integrated Sachs-Wolfe effect. 

\begin{figure}[!ht]
\hspace{1.40cm}\psfig{file=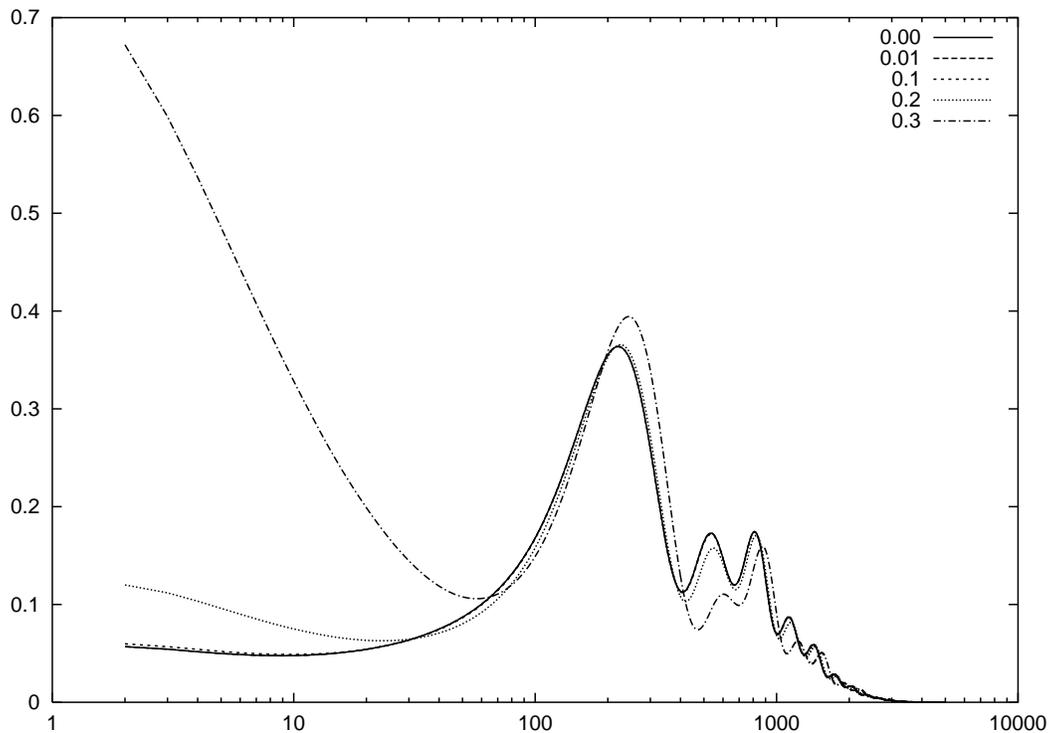,width=10.0cm,angle=270}\vspace{0.75cm}
\caption[h]{
  The temperature anisotropy power spectrum, $l(l+1)C_l/2\pi$, for the
  constant coupling case: the values in the legend are the values of
  $\alpha$.}
\label{fig:constTT}
\end{figure}

\begin{figure}[!ht]
\hspace{1.40cm}\psfig{file=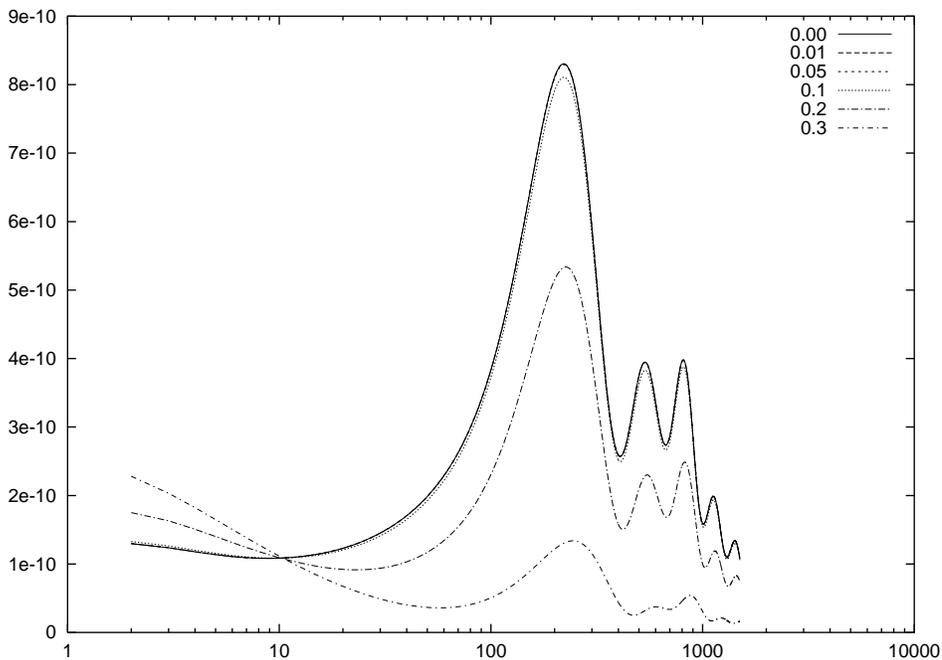,width=9cm,angle=270}
\vspace{0.75cm}
\caption[h]{
  COBE-normalized temperature anisotropy $l(l+1)C_l/2\pi$ 
  for the case of
  constant coupling. On small scales the COBE--normalized spectra are
  below the predictions for vanishing coupling due to the enhanced
  ISW.}
\label{fig:constCOBE}
\end{figure}

In Figure 2 we plot the COBE normalized curves. It can be seen that 
increasing the parameter $\alpha$ implies that there will be less power 
on small angular scales (i.e. large multipole number). Only on very 
large scales ($l\leq 10$) there is more power than in the model with 
$\alpha=0$. 

\subsection{The case of field-dependent coupling}
We now turn our attention to the field $R$, which coupling function depends on the 
field value. As already mentioned, if $R$ is zero initially, it remains zero. 
However, deep in the radiation dominated epoch, it might well be that the field 
was different from zero. 

\begin{figure}[!ht]
\hspace{1.40cm}\psfig{file=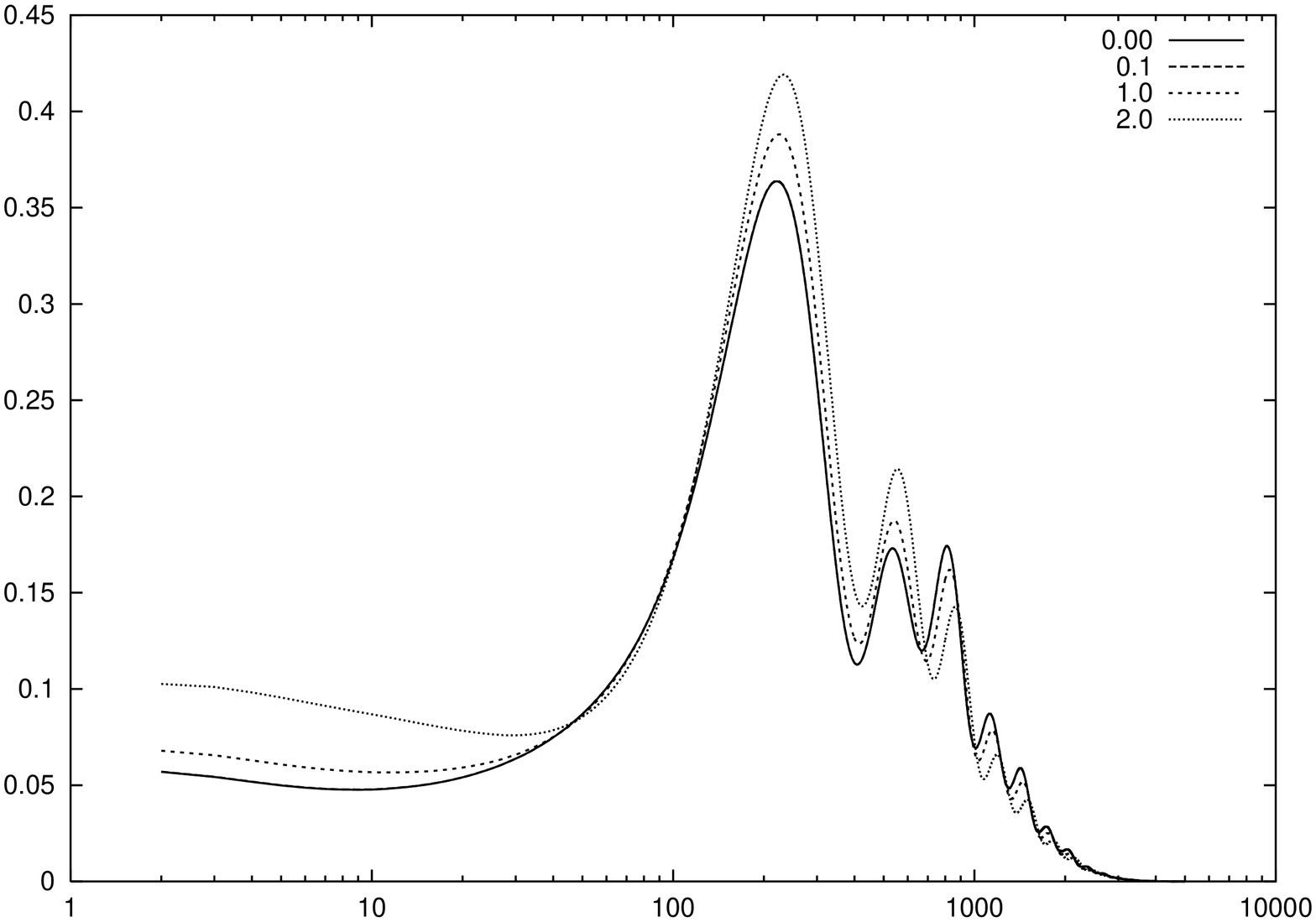,width=10.0cm,angle=270}\vspace{0.75cm}
\caption[h]{
  The temperature anisotropy power spectrum, $l(l+1)C_l/2\pi$, for the
  field-dependent coupling case: the values in the legend are for the
  initial values of the scalar field $R$.}
\label{fig:Runnormalized}
\end{figure}

In Figure 3 we plot the unnormalized power spectrum for different initial 
conditions of the field $R$. On large angular scales (low multipole number) 
one can see that the integrated Sachs--Wolfe effect is not as pronounced as 
in the case of constant coupling. This is because at low redshift ($z<1.5$) 
the field value of $R$ is quite small already, so that coupling to matter 
is small, too. We refer to \cite{rhodes} for details (see also \cite{amendola} for  
a discussion on cosmological perturbations in some classes of dilatonic 
dark energy models). 

\begin{figure}[!ht]
\hspace{1.40cm}\psfig{file=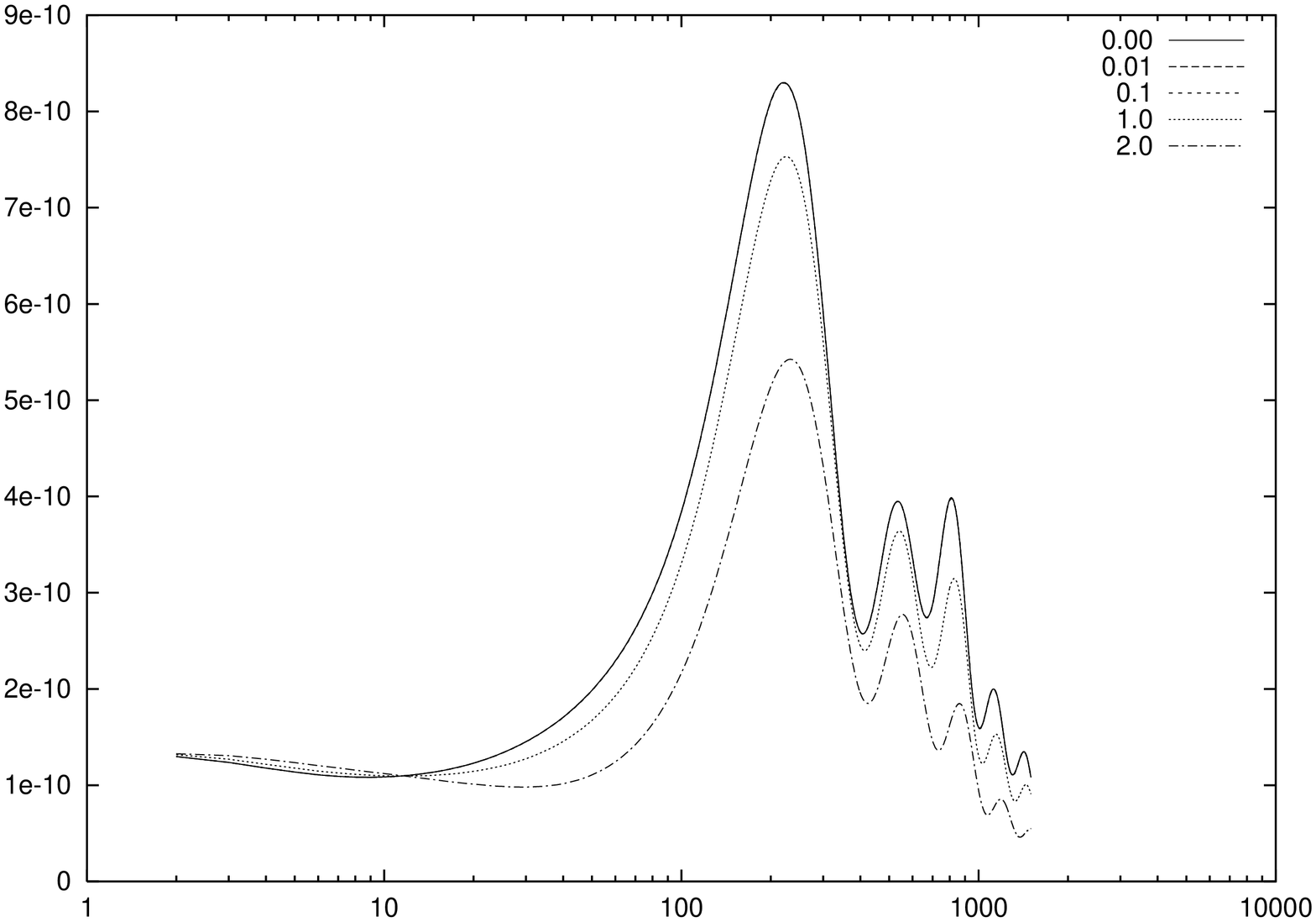,width=9cm,angle=270}\vspace{0.75cm}
\caption[h]{
  COBE-normalized temperature anisotropy $l(l+1)C_l/2\pi$ for the case of
  field-dependent coupling.  Similar to the case in figure
  \ref{fig:constCOBE}, on small scales the COBE--normalized spectra
  are below the predictions for vanishing coupling.}
\label{fig:Rnormalized}
\end{figure}

In Figure 4 we plot the COBE normalized figures. The most obvious difference 
to the case of constant coupling (Figure 2) is the region $2\leq l \leq 10$.
This is because the integrated Sachs--Wolfe effect is not as pronounced as 
in the case of constant coupling. 

\section{Conclusions}

Given that observations of anisotropies in the CMB will have very high accuracy, 
they will also provide vital constraints on models based on higher dimensions. 
In the case of brane world scenarios with bulk scalar field it can be said 
that highly warped bulk geometries are favoured (which corresponds to 
small $\alpha$, see \cite{brax}). Furthermore, the initial 
distance between the branes at the time of nucleosynthesis will be severely 
constrained by CMB experiments. 

Both moduli fields affect the CMB. However, their effect is quite different, 
due the fact that in the case of the $R$--field the coupling to matter is decreasing 
in the cosmological history. 

An interesting possibility would be to extend our analysis by giving the 
moduli fields a potential. They may then play the role of dark energy in 
the universe \cite{braxdarkenergy}. It is unlikely, however, that this can 
be done without fine tuning of some parameters in the potential. On the 
other hand, the origin of such a potential might come from a repulsion of 
the second brane from a singularity \cite{brax}. It would be interesting to 
investigate the effect of such a potential on the results presented here. 
In particular it would be important to investigate some possibilities: first, 
one could imagine a situation, where visible matter is located on our brane 
only, but dark matter only on the second brane. In this case, baryons 
and dark matter couple differently to dark energy. The second possibility 
is where dark matter and baryons live on our brane only, and the 
second brane is empty. This essentially the case studied here and in 
\cite{rhodes} for constant potential energy (i.e. cosmological constant). 
Finally, there might be two types of dark matter, one lives on our brane 
and the other dark matter type lives on the other. In this case, the 
masses of both types vary differently in time. This could imply some 
novel effects for structure formation not only on large scales. 

Another effect neglected here is the variation of the fine structure 
constant $\alpha_{\rm em}$. If $\alpha_{\rm em}$ varies, the Thomson 
cross section changes with time. Therefore the scattering between 
photons and free electrons is modified. It is known that this leaves 
an imprint in the CMB anisotropies. It might be interesting to investigate
the magnitude of this effect in the theory presented here \cite{palma}. 

In conclusion, cosmological considerations will 
provide useful complementary information to local experiments about extra 
dimensions, especially about the size of extra dimensions. 
This will not only hold for the class of brane world models 
discussed here, but also for other, maybe more exotic, models. In any case, 
cosmology continues to play a significant role in constraining theories 
beyond the standard model of particle physics. 

\vspace{1cm}

{\bf Acknowledgements:} The authors were supported in part by PPARC.

\vfill 
\end{document}